# Accuracy features for quantum process tomography using superconductor phase qubits

Yu. I. Bogdanov[1 *], S.A. Nuyanzin[1,2]

[1] *Institute of Physics and Technology, Russian Academy of Sciences, Moscow, 117218 Russia*
[2] *National Research University, Moscow Institute of Electronic Technology, Moscow, 124498 Russia*

We propose a method for precision statistical control of quantum processes based on superconductor phase qubits. Using the universal quantum tomography method, we provide a detailed analysis of accuracy of tomography for a 2-qubit gate SQiSW, which arises due to capacitive coupling between qubits. The developed approach could be successfully applied for quality and efficiency problems of superconductor quantum information technologies.

## 1. Introduction

The main obstacles to the path of building a scalable quantum computer are imperfect technologies of preparation of quantum registers, difficulties in measurements and the need to reduce decoherence. The realized accuracy (Fidelity), which is characterized by the probability of coincidence between theoretical and experimental quantum states, is about 60-80% for the latest experiments. At the same time the required accuracy must be 99.99% and higher [1,2].

One of the bottlenecks in the development of quantum information technologies is the lack of methodology to control quantum states and processes. Such methodology must play the role of the interface between development and realization of element basis of quantum computers.

In present work we propose a new method to estimate quality and efficiency of quantum gates on superconducting phase qubits. Based on the original approach developed in [3], we provide analysis for SQiSW realization (square of i-SWAP). Previously, similar research has been performed for optical qubits [4,5]. In present paper we also provide a comparison of various quantum tomography protocols, which are of practical interest for superconducting quantum information technologies.

## 2. Phase qubit

The phase qubit model is based on Josephson junction with bias current. There are many works dedicated to studies of this system (see [6-10] and refs.). Hamiltonian of the considered system in canonical form is as follows [6]:

$$H = E_C n^2 - E_J \cos\varphi - \frac{\hbar}{2e} I_e \varphi \qquad (1)$$

Where $\varphi$ - is the phase of junction, $n$ - is the number of Cooper pairs

Here $E_C = \frac{(2e)^2}{2C}$ - is the electrostatic energy parameter coupled with junction capacity, $E_J = \frac{\hbar}{2e} I_c$ - is the Josephson energy that is proportional to critical current $I_c$.

A phase qubit corresponds to quantum oscillations of a phase particle near the bottom of the effective potential. The considered mode works in case of $E_C << E_J$. The potential minimum corresponds to $\varphi_0$, where $\partial U / \partial \varphi = 0$. This point is defined with by condition: $I_e = I_c \sin\varphi_0$

Oscillation frequency near the bottom (called plasma frequency) is

$$\omega_p = \omega_J \left(1 - \left(\frac{I_e}{I_c}\right)^2\right)^{1/4}, \text{ where } \omega_J = \sqrt{\frac{2eI_c}{\hbar C}} \text{ -is the Josephson frequency}$$

---
[*] E-mail: bogdanov@ftian.ru

The state energy spectrum in harmonic approximation is

$$E_k = \hbar \omega_p (k + 1/2), \text{ where } k = 0, 1, \ldots \qquad (2)$$

Note that because of the potential anharmonicity the energy level spectrum is not strictly equidistant. It is important from the technology point of view because it allows one to select two bottom levels to create phase qubit logical states $|0\rangle$ and $|1\rangle$ (see Figure 1).

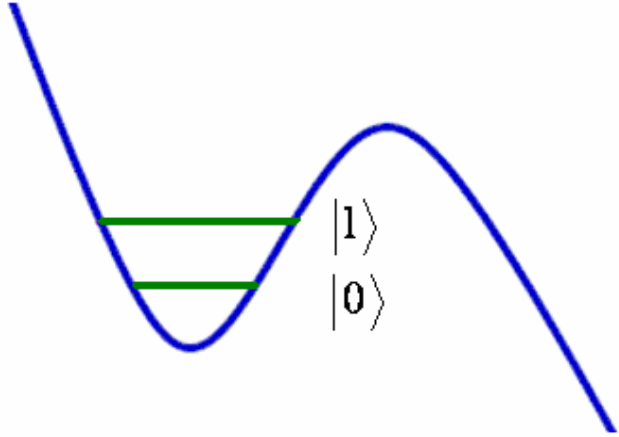

Fig.1 Schematic depiction of basis states of phase qubit in effective potential well.

Strictly speaking, there are no bound states in the considered well. The considered qubit states are metastable and they are vulnerable to macroscopic quantum tunneling, though one make the corresponding lifetimes very large by tuning operational parameters

Let the ground state correspond to the logical zero of a qubit $|0\rangle = \begin{pmatrix} 1 \\ 0 \end{pmatrix}$, and the first excited state - to the logical one of the qubit $|1\rangle = \begin{pmatrix} 0 \\ 1 \end{pmatrix}$. Hence the qubit Hamiltonian could be written as

$$H_{qubit} = -\frac{1}{2}\varepsilon \sigma_z, \text{ where } \varepsilon = E_1 - E_0, \quad \sigma_z = \begin{pmatrix} 1 & 0 \\ 0 & -1 \end{pmatrix}$$

Here the state energy is going to be $-\varepsilon/2$ for the ground and $+\varepsilon/2$ for the excited states respectively.

Quantum computations relate to execution of a sequence of unitary transformation by using quantum logical elements (gates) to qubit registry. At the end of this sequence, there is a quantum measurement performed upon qubits (generally speaking destructive). Let's assume that before measurement there were both top and bottom energy states of the qubit presented in the superposition. To determine the probability of the top state one can slowly increase the bias current until the top energy level reaches the top of the potential barrier, which will make the phase particle tunnel fast from this level. Then if the qubit after state measurement is at the top level, the Josephson junction will switch from the Josephson branch to dissipative branch. That could be registered by the measurement of output average voltage across junction (the junction would be in a non-zero voltage state). If the qubit is in the lower energy state then the tunneling probability will be small, and the junction will stay at Josephson branch resulting in zero output voltage. The alternative method to activate the switch is based on applying a high-frequency signal with resonance frequency (instead of control of potential). In this case the top energy level will be excited, which will lead to the switch of the junction. This approach is quite similar to the standard method of reading states in atom physics.

The two-qubit transformation SQiSW that we consider in this work, is performed as the result of capacity coupling of superconducting qubits [7,8]

$$H_{int} = \hbar(g/2)(|01\rangle\langle 10| + |10\rangle\langle 01|), \quad (3)$$

where $g$ - is the interaction constant.

The considered Hamiltonian defines evolution by the next following matrix:

$$U_{int} = \begin{bmatrix} 1 & 0 & 0 & 0 \\ 0 & \cos(gt/2) & -i\sin(gt/2) & 0 \\ 0 & -i\sin(gt/2) & \cos(gt/2) & 0 \\ 0 & 0 & 0 & 1 \end{bmatrix} \quad (4)$$

Pi-impulse, that satisfies condition $gt = \pi$, guarantees the swap (i-SWAP): $|01\rangle \to -i|10\rangle$, $|10\rangle \to -i|01\rangle$. The impulse of half the duration i.e. $gt = \pi/2$ provides the required operation SQiSW.

Note that the important quantum informatics transformation CNOT could be realized based on SQiSW and single-qubit rotations:

$$CNOT = R_y(-\pi/2) \otimes I \cdot R_x(\pi/2) \otimes R_x(-\pi/2) \cdot SQiSW \cdot R_x(\pi) \otimes I \cdot SQiSW \cdot R_y(\pi/2) \otimes I$$

Here $I = \begin{pmatrix} 1 & 0 \\ 0 & 1 \end{pmatrix}$ - is the identity qubit transformation, $R_x(\alpha)$ $R_y(\alpha)$ - are Bloch sphere operators' of rotation by specified angle relative to axes $x$ and $y$.

## 3. Quantum operations and quantum tomography

An ideal quantum logical gate provides unitary transformation for a quantum state (density matrix): $\rho_{out} = U\rho_{in}U^+$.

The real state evolution could never be unitary. One has to account for inevitable interaction between quantum system and environment (quantum noise). In open quantum system theory the evolution is defined by the sum of operators $\mathsf{E}(\rho) = \sum_k E_k \rho E_k^+$ [1,11]. For this reason the coupling between input and output states is defined with formula $\rho_{out} = \sum_k E_k \rho_{in} E_k^+$, where $E_k$ - are so-called transformation elements (Kraus operators), which satisfy the following normalization condition (for transformations that preserve trace): $\sum_k E_k^+ E_k = I$, where $I$ -is the identity matrix.

The representation of a quantum process as a sum of operators guarantees that if there is a Hermitian positively defined matrix of trace one at the input (i.e. the density matrix), then there will also be a Hermitian positively defined matrix of trace one at the output. It follows that the representation as the operator sum guarantees not only positiveness but also so called complete positiveness (the appropriate mapping is completely positive) [1,11].

Operators $E_k$ could be represented by matrices with size $s \times s$ in Hilbert space of dimension $s$. In the case of a unitary transformation, there is only one summand defined by $U$ operator in the operator sum. The arbitrary transformation could be reduced to the form that contains no more than $s^2$ matrices of $E_k$.

Based on transformation elements one could build a so-called chi-matrix, which plays a key role in tomography of quantum processes. Let's take matrix $E_1$ with size $s \times s$ and stretch it into a column $e_1$ of length $s^2$ (so the second column shall be under the first one and so on). The obtained column $e_1$ will be the first column of some matrix $e$. Similarly, matrix $E_2$ will define the second column of $e$ and so on. Based on the obtained matrix $e$ we shall define the main matrix $\chi$ of dimension $s^2 \times s^2$ as $\chi = ee^+$

That's important to note that matrix $\chi$ could be interpreted as a matrix in Hilbert space of dimension $s^2$. So any quantum transformation in Hilbert space of dimension $s$ is reduced to a state in Hilbert space of dimension $s^2$ - so-called Choi-Jamiolkowski isomorphism [11]. The corresponding state could be considered as a joint state of two subsystems A and B (each of dimension $s$). Due to normalization condition (trace preservation) the reduced matrix $\chi_A$ that is derived when taking trace by subsystem B must be proportional to identity matrix with size $s \times s$: $\chi_A = Tr_B(\chi) = const \cdot I$

Matrix $e$ and, therefore, matrices $E_k$ are defined ambiguously. Let matrix $e$ have $m$ columns and size $s^2 \times m$ (hence there are $m$ transformation elements $E_k$, $k=1,2,...,m$). The chi-matrix will not change after the following transformation: $e \to e' = eU$, where $U$ - is the unitary matrix with size $m \times m$. There will be new matrices $E_k'$, which are unitary equivalent to the set of initial matrices $E_k$ and that correspond to the new matrix $e'$.

On the other hand, the form of the chi-matrix depends on representation. Let $|j\rangle$ be a ket-vector (column), where the $j$-th element is equal to one and other elements are zeros. Similarly, $\langle k|$ -is a bra vector (raw, where the $k$-th element is equal to one and other elements are zeros). Let us consider matrix $|j\rangle\langle k|$, where the element at crossing of the $j$- th raw and the $k$- th column is equal to one and other elements are zeros. Let indexes $j$ and $k$ take values from 1 to $s^2$ ($j,k=1,2,...,s^2$). There are $s^4$ such matrices. Evidently, the chi-matrix can be written as the following decomposition: $\chi = \sum_{j,k} \chi_{jk} |j\rangle\langle k|$.

In this case the set of $s^4$ matrices $|j\rangle\langle k|$ plays the role of basis. When moving from matrices $|j\rangle\langle k|$ to other base sets of matrices the decomposition coefficients change will change of course. It corresponds to the transition to another chi-matrix representation. Following Chuang and Nielsen [1,12], let us consider a single-qubit base set defined by the following four matrices 2x2:

$$I_0 = I/\sqrt{2}, \ X = \sigma_x/\sqrt{2}, \ Y = -i\sigma_y/\sqrt{2}, \ Z = \sigma_z/\sqrt{2} \qquad (5)$$

These matrices are built from the identity matrix and Pauli matrices.

Then the two-qubit base set will be constructed from tensor products of various pairs of considered matrices (16 operators in total). Similarly, the three-qubit base set will take tensor products of various triads of considered matrices (64 operators in total) and so on.

Note that there is always a unitary matrix $U_0$, which provides transition to chi-matrix presentation in a new basis: $\chi' = U_0^+ \chi U_0$

## 4. Results of modeling

In this section, we shall demonstrate the advantages of our approach compared with the traditional one. The traditional approach developed in [1,12], has been applied to tomography of superconducting phase qubits in novel experiments by the group of Prof. Martinis [7,8]. It is based on tomography measurements using the following set of 4 input single-qubit states:

$$|0\rangle, |1\rangle, \frac{1}{\sqrt{2}}(|0\rangle + |1\rangle), \frac{1}{\sqrt{2}}(|0\rangle + i|1\rangle) \qquad (6)$$

The respective two-qubit input states are formed based on tensor products of single-qubit states (16 states in total).

It appears that a more optimal set of single-qubit states is the one proposed in [13]. The set is based on symmetry of regular tetrahedron. Let imagine a regular tetrahedron inscribed into Bloch's sphere. The appropriate states are produced by vectors going from the centre perpendicularly to the faces of the tetrahedron. Similarly, we may consider basis sets based on symmetry of other regular, as well as non-regular but highly symmetrical polyhedrons [14].

To compare different protocols one may use the universal statistical distribution of quantum tomography accuracy proposed in [3]. There the accuracy of reconstruction of chi-matrix, which is analogous to the density matrix for quantum tomography, is considered. The respective fidelity $F$ is characterized by the probability of coincidence between the experimentally reconstructed chi-matrix $\chi$ and the theoretical chi-matrix $\chi_0$:

$$F = \left( Tr \sqrt{\chi_0^{1/2} \chi \chi_0^{1/2}} \right)^2 \qquad (7)$$

In this equation it is implied that the chi-matrix trace is normalized to one.

The value $dF = 1 - F$ can be naturally called the loss of fidelity. The loss of fidelity is a random variable and its asymptotic distribution could be written as [3]:

$$1 - F = \sum_{j=1}^{j_{max}} d_j \xi_j^2 \qquad (8),$$

where $d_j \geq 0$ are non-negative coefficients, $\xi_j \sim N(0,1)$, $j = 1,...,j_{max}$ are independent normally distributed random variables with zero mean and variance equal to one.

As an example of application of our approach, on Fig.2 we present the accuracy of tomography of quantum process defined with SQiSW in depolarizing quantum noise. The respective quantum transformation affects the density matrix $\rho$ of two-qubit state as:

$$\rho \rightarrow \frac{pI}{4} + (1-p) U \rho U^+ \qquad (9)$$

Here $I$ is an identical matrix with size 4x4. The replacement of the initial state with an absolutely chaotic state (white noise) is realized with probability $p$ and the unitary transformation by matrix $U$ is realized with probability $1-p$. In the considered example $p = 0.5$. The top figure shows density distributions for fidelity loss $dF = 1 - p$. The distributions are multiparametric (240 parameters). The solid curve in the figure corresponds to the standard set of inputs, the dashed line to the optimized one. The sample size is 1 million.

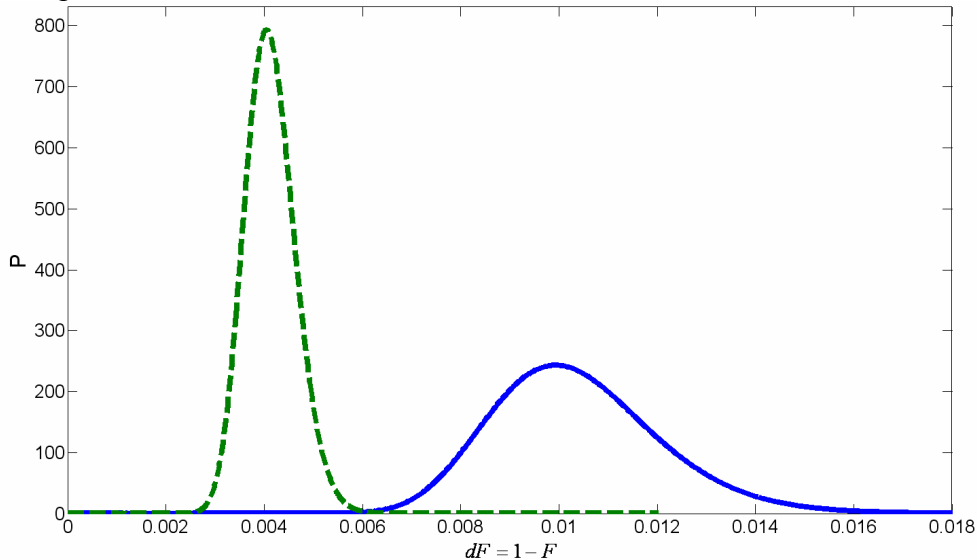

Fig. 2 Distribution of loss of fidelity for quantum tomography of gate SQiSW that is affected by depolarizing noise with weight p=0.5. Solid line corresponds to standard basis sets, dashed line – basis set on tetrahedron symmetry.

As is evident from the figure, the optimization allows one to decrease average fidelity loss by 2.5 times. Fig. 3 demonstrates the real and imaginary parts of the chi-matrix. The basis made of all possible tensor products of matrices (5) is used.

It was already mentioned above that the considered gate SQiSW defines a natural two-qubit transformation for superconducting phase qubits [7,8]. Gates CNOT and CZ could be constructed based on SQiSW and single-qubit transformations. It can be demonstrated that the results for these gates are absolutely analogous to those presented on Fig. 2.

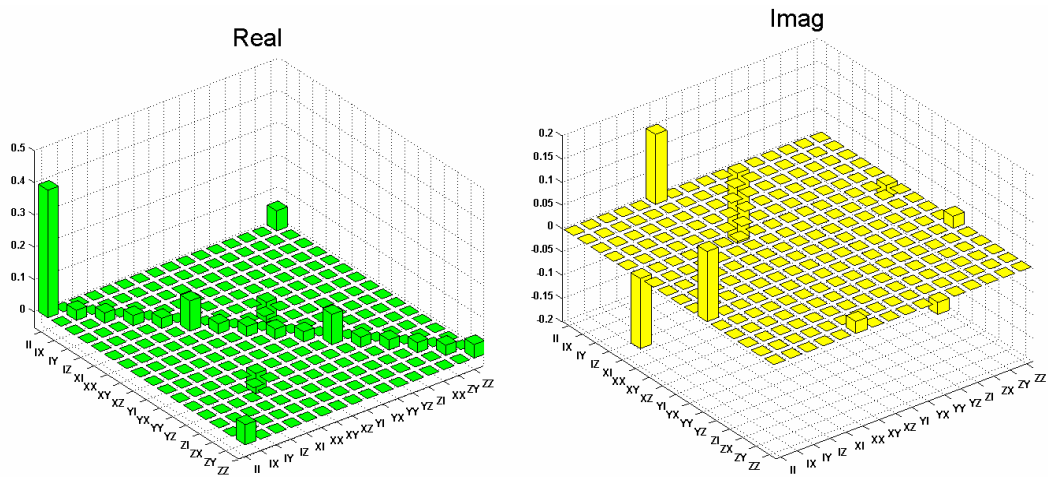

Fig.3. Visualization of chi-matrix for considered gate. Left – real part, right – imaginary part.

Let us note some important advantages of the offered approach compare to the standard one.

We realize a direct statistical estimation of chi-matrix by experimental data in accordance with the approach developed in [3]. At the same time the two-stage procedure proposed in [1,12] and implemented in [7,8] is not quite correct. Our approach ensures a correct registration of statistical fluctuations and instrumental errors in experiments and technology. Therefore a throughout assessment of completeness, adequacy and accuracy of quantum measurement protocols is provided.

Note that the issues considered in present work are left out of view in other studies. As the result one can not adequately assess which precision could be obtained in every measurement protocol, how one can improve the fidelity of control, how can adequacy of measurement results be analyzed etc.

## 5. Summary

In present work we develop a method of precision control of quantum processes tomography, based on superconductive phase qubits.

Based on the universal approach developed earlier by one of the authors, we conduct a throughout analysis of accuracy of quantum tomography based on two-qubit gate SQiSW. This approach can be successfully applied to analysis of various quantum processes based on superconducting quantum information technologies.

We demonstrate that one can significantly increase the accuracy of quantum tomography compared to the traditional approach by using an input set of quantum states corresponding to the symmetry of a regular tetrahedron.

Finally, the developed method provides a complete picture with regards to quality and efficiency of considered quantum registers. It allows one to formulate the specifications for experimental apparatus and requirements to technologies. Thus, the feedback from the approach allows one to make best use of available resources and optimize the process of development quantum information technologies.

We would like to thank Vladimir F. Lukichev and Alexander A. Orlikovsky for discussions of work results.

The work has been supported by Program of Russian Academy of Sciences in fundamental research.